\nonstopmode

\documentclass[aps,prl,twocolumn,showpacs]{revtex4}
\usepackage{graphicx}
\begin{document}
\title{Magneto-transport of large CVD-grown graphene}
\author{E. Whiteway}
\author{V. Yu}
\author{J. Lefebvre}
\author{R. Gagnon}
\author{M. Hilke}
\affiliation{Department of Physics, McGill University, Montr\'eal, Canada H3A 2T8}
\date{\today}

\begin{abstract}
We present magnetoresistance measurements on large scale monolayer graphene grown by chemical vapor deposition (CVD) on copper. The graphene layer was transferred onto SiO$_2$/Si via PMMA and thermal release tape for transport measurements. The resulting centimeter-sized graphene samples were measured at temperatures down to 30mK in a magnetic field. We observe a very sharp peak in resistance at zero field, which is well fitted by weak localization theory as well as strong localization. The samples exhibit conductance fluctuations symmetric in field, which are due to large scale inhomogeneities consistent with the grain boundaries of copper during the CVD growth.

\end{abstract}
\maketitle

Graphene has attracted a considerable amount of attention due to the ease in isolating a single sheet of graphite via mechanical exfoliation \cite{novo04,novo05}. Despite the fact that it is one atom thick, exfoliated graphene has shown extraordinary transport properties and can be used as a novel material for many potential applications. Its unique band structure has led to many interesting phenomena such as tunable charge carriers between electrons and holes, anomalous integer quantum Hall effect \cite{novo04,zhang05} and ultrahigh mobilities at room temperature \cite{bol08}. The most noteworthy property of the band structure is the existence of two degenerate Dirac cones \cite{Wallace47}, which leads to two degenerate valleys (K and K').

In a seminal work by McCann and co-workers \cite{WLth} based on earlier work on honeycomb lattices \cite{ando02}, the authors have obtained a general expression for the weak localization (WL) correction in graphene, which determines the dependence of the magnetoresistivity as a function of the magnetic field (B) for various scattering parameters in graphene. The increase in resistance at zero field is due to an increased backscattering probability when a path and its time reversed path is phase coherent. This time reversal symmetry is destroyed by a small B, which leads to a decrease in resistance with field. This generic WL effect is suppressed by inelastic scattering due to a loss of phase coherence. In graphene, however, backscattering can only occur when scattering between the two valleys is possible (inter-valley scattering). This depends on the type of scattering potential. Typically, slow varying potentials will lead to no inter-valley scattering, whereas short range impurities will. The existence of inter-valley scattering and intra-valley scattering (within a valley) leads to an interesting competition between WL and weak anti-localization (WAL), since in the absence of inter-valley scattering, only WAL exists, and is responsible for the increase in resistance with B. Intentionally disordered exfoliated graphene promotes inter-valley scattering, which can then lead to strong localization \cite{moser10}.

While early experiments on exfoliated graphene have shown a strong suppression of WL even at very low temperatures \cite{morozov08}, more recent experiments have shown significant WL effects by averaging over many carrier concentrations \cite{tikh08,ki08,mason10}. This was necessary because mesoscopic conductance fluctuations dominate transport in these small graphene flakes since the sample sizes are comparable to the phase coherence length $L_\phi\simeq 1\mu$m at low temperatures. Transport in large scale graphene would be self-averaging when the sample size is much larger than $L_\phi$. This is what we propose to demonstrate in this letter.

While acquiring large graphene flakes by exfoliation might not be feasible, other techniques have been developed for large-scale production, such as epitaxial graphene grown on a SiC(0001) surface, where WAL has been observed \cite{wu07}. Recently, nickel films \cite{reina08} and copper foils \cite{li09} have been successfully used as substrates to produce large scale single layer graphene by chemical vapor deposition (CVD).

For this work we grew graphene monolayers by CVD of hydrocarbons on 25 $\mu$m-thick commercial Cu foils. The Cu foil is first acid-treated for 10 mins using acetic acid and then washed thoroughly with de-ionized water. Graphene growth is realized in conditions similar to Li et al. \cite{li09}, but using a vertical quartz tube. Graphene is grown at 1025$^\circ$C in 0.5 Torr, with a 4 sccm H$_2$ flow and a 40 sccm CH$_4$ flow for 30 minutes. The methane flow is stopped while the hydrogen flow is kept on during the cooling process.

\begin{figure}[ptb]
\begin{center}
\vspace{2cm}
\includegraphics[width=3.5in]{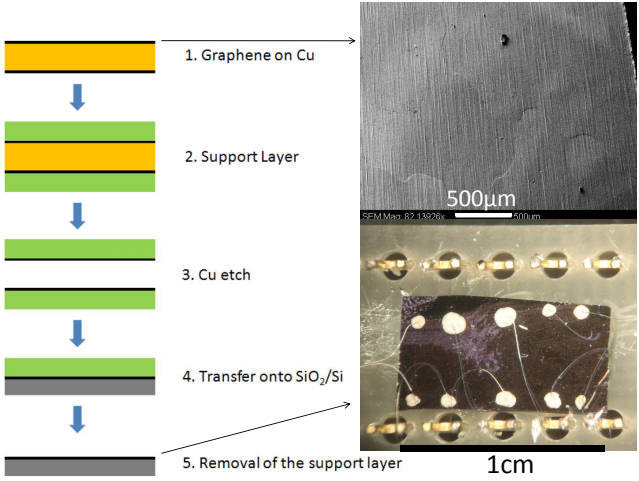}
\vspace{-2cm}
\caption{Left: transfer process of the graphene monolayer. Top right: SEM picture of graphene on copper showing typical grain boundaries. Bottom right: picture of the 1$\times$0.5cm$^2$ graphene device on a SiO$_2$/Si substrate used in this work with mm-sized silver paste contacts.}%
\label{fabrication}%
\end{center}
\end{figure}

The graphene film was transferred onto n$^{++}$-doped Si wafer with an oxide thickness of 285nm. This is realized by etching the Cu foil in an oxidizing solution of 0.1M ammonium persulfate ((NH$_4$)$_2$S$_2$O$_8$). PMMA or Nitto Denko thermal release tape (3195MS) are used as a supporting layer. For some devices, we did not remove the PMMA layer unlike Li {\it et al.}\cite{li091} but used it to provide an additional insulation between the gate and the graphene as well as to keep the integrity of the film. As for the thermal release tape, its adhesive force is removed by heating the tape to 120$^\circ$C in order to detach the graphene film and deposit it onto SiO$_2$/Si. Small contacts of SPI conductive silver paint, arranged according to the van der Pauw technique, are placed on the periphery of the graphene film and on the Si backgate. Silver wires are used to connect the contacts to the chip carrier to perform transport measurements. An image of the sample is shown in fig. \ref{fabrication}. Dozen cm$^2$-sized samples were made and all were conducting, with four terminal resistivities ranging between 1 and 4k$\Omega$ at room temperature. All samples show a weak increase in resistance upon cooling to 4.2K. Contact resistances vary between 1 and 10k$\Omega$. Raman spectroscopy and optical reflection were used to confirm the monolayer nature of graphene on the devices \cite{Yu09}.

\begin{figure}[ptb]
\begin{center}
\includegraphics[width=3.5in]{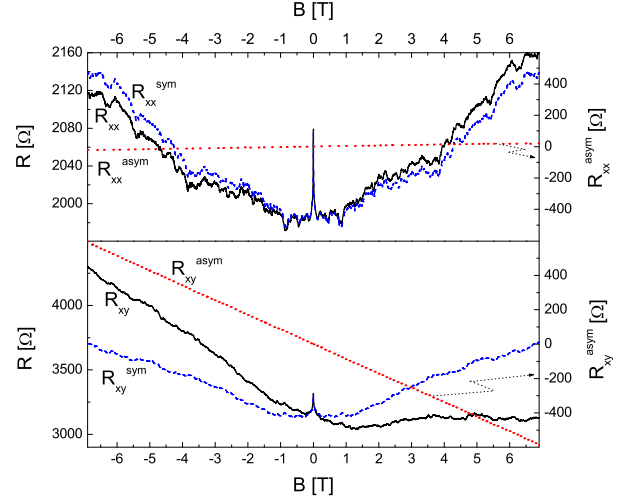}
\caption{Top graph: Magnetoresistance ($R_{xx}$) as a function of B at 30mK. Also shown are the extracted symmetric ($R_{xx}^{sym}=[R_{xx}(B)+R_{xx}(-B)]/2$) and antisymmetric ($R_{xx}^{asym}=[R_{xx}(B)-R_{xx}(-B)]/2$) traces. Bottom graph: Hall resistance ($R_{xy}$) as a function of B and the corresponding symmetric and antisymmetric components. The Hall slope is $R_{xy}^{asym}$=-(85.29 $\pm$ 0.005)$\Omega$/T.}
\label{sym}%
\end{center}
\end{figure}

The samples were placed inside a dilution refrigerator and measured as a function of B. A most striking feature is a sharp peak of the resistance at $B=0$ as shown in fig. \ref{sym}. Similar peaks were obtained in all samples measured.

Because we have no lithographically defined contacts, which avoids possible residues from lithography, there is a large symmetric contribution in the Hall trace due to contact misalignment. However, this large symmetric signal stemming from the relatively large magnetoresistance ($\sim$2k$\Omega$) can be effectively eliminated by antisymmetrization. The result is a remarkably linear antisymmetric Hall trace (see fig. \ref{sym}), which shows no indication of quantum Hall plateaus. The slope can be used in order to effectively extract the total Hall carrier density, i.e., $n=(e\partial R_{xy}^{asym}/\partial B)^{-1}\simeq 7.3\cdot 10^{12}$cm$^{-2}$. By applying a gate voltage we observe a change in the Hall slope (density) of the sample. In the remainder we will use a zero gate voltage configuration for a more detailed analysis. This (anti-)symmetrization method described above can also be applied to the magnetoresistance trace in fig. \ref{sym}, where we observe a very small linear antisymmetric contribution. This small antisymmetric Hall contribution in $R_{xx}$ comes from a non-ideal contact region. In most cases this contribution can be neglected.

\begin{figure}[ptb]
\begin{center}
\includegraphics[width=3.5in]{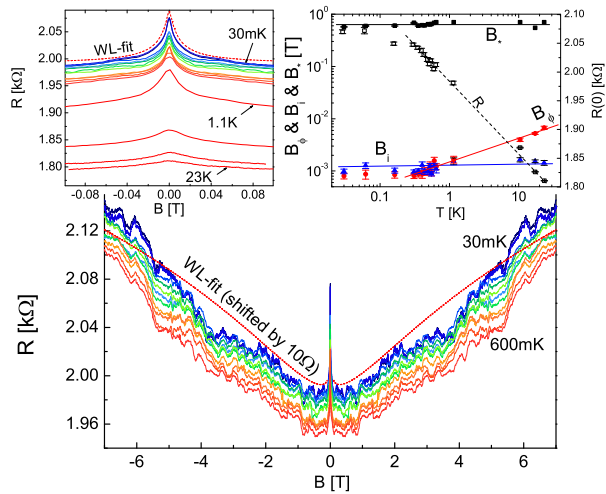}
\caption{In the bottom graph we show the symmetric resistance as a function of B for different temperatures (30-600mK). The WL fit is shown with a dashed line and offset by +10$\Omega$ for clarity and without the $B\simeq 0$ region. The top left graph shows the same data plus temperatures from 1.1 to 23K but zoomed in around zero field. The top right graph shows the temperature dependence of the results to the best fits from equation (1) for parameters $B_\phi$, $B_i$, $B_\star$ and $R(0)$ with lines as guides for the eye. The lowest two temperatures were measured one week apart from the other temperatures and show very small changes in the pattern of the resistance fluctuations.}
\label{WLfit}%
\end{center}
\end{figure}

We now turn to understanding the large and very narrow peak at $B=0$, which strongly suggests WL. Following the theoretical work by McCann et al. \cite{WLth}, we can write the magnetoresistivity as a function of B as:

\begin{equation}
\rho(B)=\rho(0)-\frac{\rho^2}{\pi \rho_0}(
F(B_\phi)-F(B_\phi+2B_i)-2F(B_\phi+B_i+B_\star)),
\end{equation}

where $F(z)=\psi(0.5+z/B)-\ln(z/B)$, $\psi$ is the digamma function and $\rho_0=h/e^2$ is the quantum unit of resistance. $B_\sigma$ with $\sigma=\phi$, $i$, or $\star$ is the characteristic B associated with the scattering time $\tau_\sigma=e\sqrt{\pi n}\rho(0)/2\pi v_FB_\sigma$, where $v_F\simeq 10^6$m/s is the graphene Fermi velocity. $\tau_\phi$ is the dephasing time, $\tau_i$ the inter-valley scattering time and $\tau_\star$ the intra-valley scattering time. We can now use equation (1) in order to fit our experimental data. The result is shown in fig. \ref{WLfit}.

Overall, the quality of the fit is quite remarkable considering that the data is obtained without performing any averaging unlike the case for exfoliated graphene \cite{tikh08,ki08,mason10}. While $B_\phi$ is strongly determined by the curvature very close to $B=0$, the fitting parameters $B_i$ and $B_\star$ are less well defined and strongly depend on the range of fields over which the fit is performed. Therefore, we first perform a fit over a wide range in B (-6 to 6T), which strongly determines $B_\star$ and then fit for $B_\phi$ and $B_i$ over a much narrower range (-.08 to .08T) using $B_\star$ obtained from the first fit. The temperature dependence of the parameters is shown in fig. \ref{WLfit}. An important additional parameter in the theory is the value of the resistivity at zero field $\rho(0)$. In particular, $\tau_i$ is strongly dependent on the value of $\rho(0)$. We illustrate this in table 1, where we show the values of various parameters for $\rho(0)=1$k$\Omega$ and $\rho(0)=3$k$\Omega$.

\begin{table}
\begin{tabular}{|c|ccccccccc|}
\hline
$\rho$ [$k$$\Omega$] & $\mu$ & $\tau_{tr}$ [fs] & $\tau_{\star}$ [fs] & $\tau_{\phi}$ [ps] & $\tau_{i}$ [ps] & L$_{tr}$ & L$_{\star}$ & L$_{\phi}$ & L$_{i}$\\
\hline
1 & 850 & 27 & 56 & 61 & 2.2 & 27 & 27 & 910 & 170 \\
3 & 280 & 9 & 61 & 46 & 37 & 9 & 17 &  450 &  410 \\
\hline
\end{tabular}
\caption{Values of various parameters for the lowest temperature data (30mK) for two values of the resistivity. The mobilities $\mu$ are in cm$^2$/V$\cdot$s and the scattering lengths L$_\sigma$ in nm. The geometric factor of the sample gives a resistivity of $\rho$=(3$\pm$0.5)k$\Omega$ and the fits are good over a wider range (up to 6T) than for a resistivity of 1k$\Omega$ (up to 4T). 95\% confidence intervals for fits of the scattering times are similar for the two resistivities and are shown in figure \ref{WLfit}.}
\label{table}
\end{table}

The mobility is given by $\mu=1/en\rho(0)$, the transport time by $\tau_{tr}=\rho_0/2\rho(0)\sqrt{\pi n}v_F$ \cite{WLth}, the mean free path by $L_{tr}=v_F\tau_{tr}$ and the other scattering lengths by $L_\sigma=v_F\sqrt{\tau_\sigma\tau_{tr}/2}$ ($\sigma=\phi,i,$ or $\star$). While the physical dimension of the contact configuration on the sample leads to a resistivity of $\sim$ 3k$\Omega$, it could well be that the actual relevant resistivity is significantly smaller if the sample is inhomogeneous. For instance, $\rho(0)$=1k$\Omega$ corresponds to the lower bound from the different samples we measured. In the presence of large inhomogeneities, the current will meander along the current path with the least resistance and hence the actual current path might be significantly longer than the direct one, which would reduce the effective resistivity compared to the geometrically defined one. Hence the values of $B_i$ and $\tau_i$ have to be taken with some caution as they are very dependent on the value of resistivity. For instance, a change of factor 3 in resistivity can affect $\tau_i$ by more than a factor 10 because the best fits for $B_i$ give a difference of a factor 20. Interestingly, when only considering scattering lengths, differences are not as marked (factor 2.5) as shown in table 1.

In general, $B_i$ determines the field at which the negative magnetoresistance due to WL stops and turns over to WAL. If $B_i=0$ there is no WL and only WAL is visible, whereas for $B_i\rightarrow\infty$ only WL survives. Hence the ratio $B_i/B_\phi$ sensitively determines the height of the WL peak. It is interesting to compare our scattering lengths with those obtained for exfoliated graphene. For instance, experiments on exfoliated graphene have obtained values for $L_i$ ranging between 75 and 250nm \cite{tikh08,ki08,mason10}, whereas for the maximum $L_\phi$ these experiments obtained values ranging between 0.6 and 1$\mu$m and for $L_\star$ values between 20 and 100nm. Overall, the values are quite similar and of the same order, with $L_\phi>L_i>L_\star$ at the lowest temperatures. With increasing temperature, only $L_\phi$ decreases due to increased inelastic scattering, while the other scattering lengths remain largely temperature independent. There is also a monotonous decrease in resistance with increasing temperature, which cannot be explained by WL alone, since WL would predict a much smaller effect \cite{WLth}.

\begin{figure}[ptb]
\begin{center}
\includegraphics[width=3.5in]{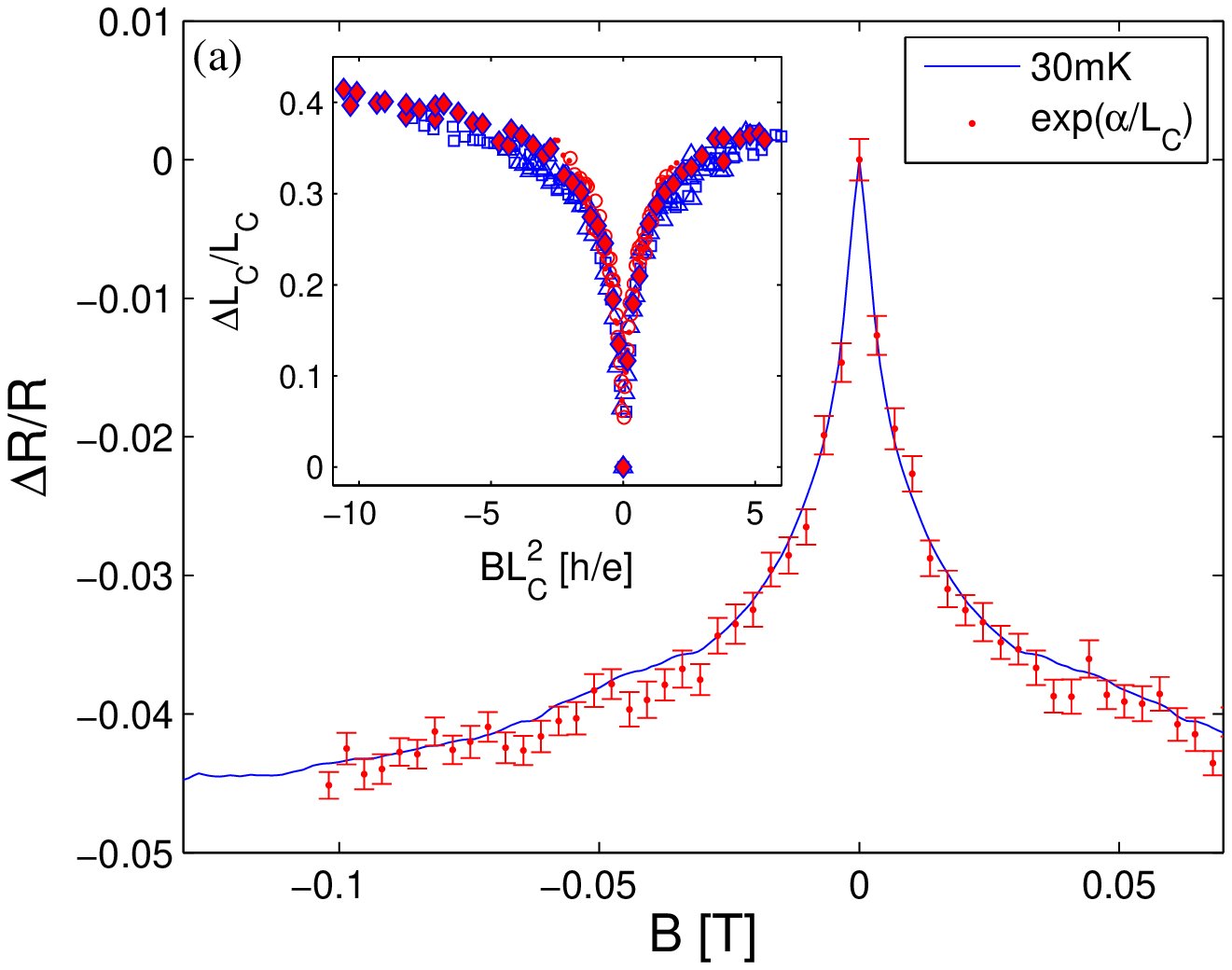}
\includegraphics[width=3.0in]{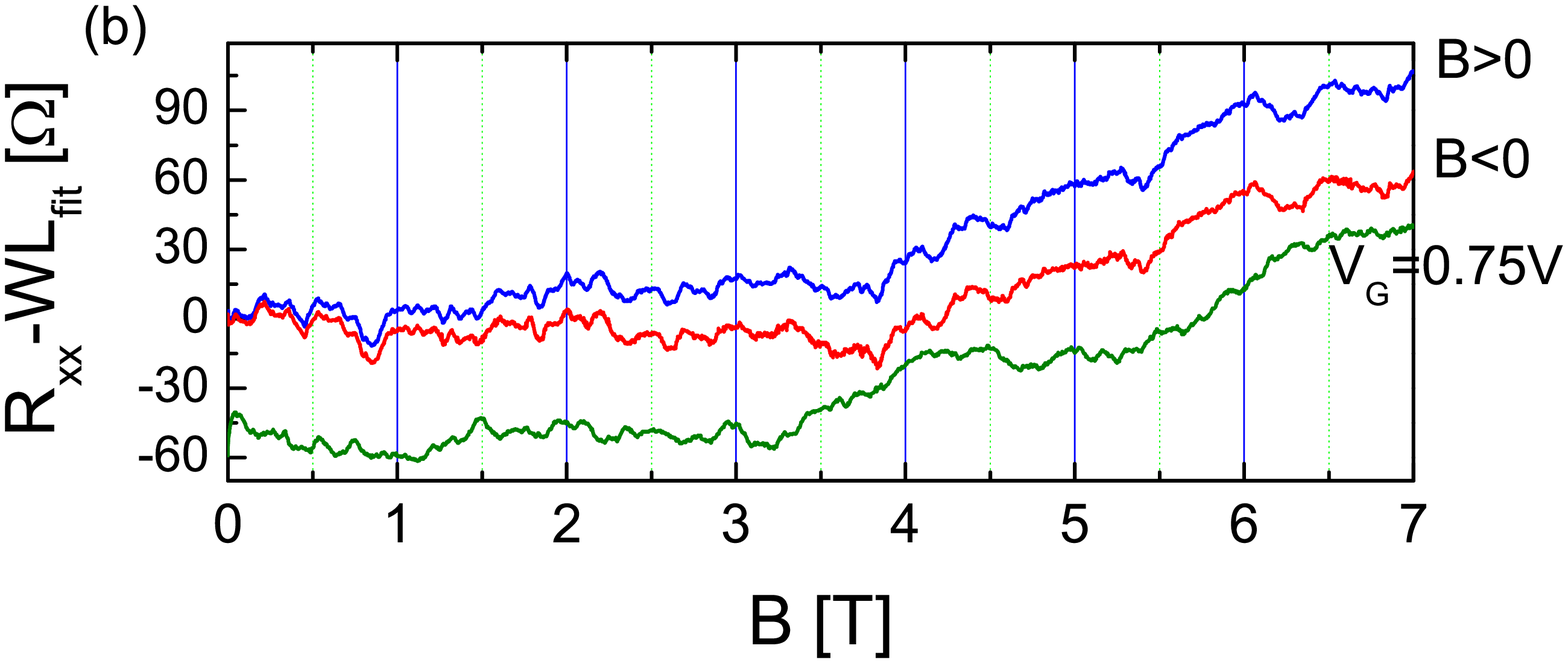}
\caption{(a) Inset: $\Delta L_c/L_c$ as a function of the magnetic flux. The different data points correspond to different values of disorder and sample widths (W=40-160 in units of carbon atoms), which all collapse on the same curve. Main figure: $\Delta \rho/\rho$ as a function of $B$ for the 30mK data (straight line) and for $\rho\sim \exp(\alpha/L_c)$ (points) with $\alpha/L_c(B=0)\simeq 0.11$. The error bars correspond to the standard error of the mean obtained from the numerical calculation of $L_c$ for graphene ribbons of widths W. (b) Resistance fluctuations as a function of the magnetic field shown as the difference from the WL fit.}
\label{fft}%
\end{center}
\end{figure}

It is therefore instructive to analyze the resistance peak at $B=0$ in terms of strong localization. Indeed, the localization length, $L_c$, in graphene is determined by a ``universal" B-dependence (independent on disorder and size) of $\Delta L_c(\phi)/L_c(\phi)$ as shown in the inset of figure \ref{fft}. $\Delta L_c(\phi)=L_c(\phi)-L_c(0)$ and $\phi=B\cdot\min\{L_c^2,L_\phi^2\}$ is the amount of flux within the area determined by $L_c^2$ or $L_\phi^2$. For $L_\phi\ll L_c$ and assuming a localized behavior $\rho\sim e^{L_\phi/L_c}\simeq 1+L_\phi/L_c$, the dip in the localization length at $B=0$ is given by $\Delta L_c/L_c(B)\simeq -0.4$ (see fig. \ref{fft}), which leads to a peak in the resistivity given by $\Delta\rho/\rho(0)\simeq 0.4L_\phi/L_c(0)$. The 30mK data can then be nicely fitted with this expression and yields $L_\phi/L_c(0)\simeq 0.11$ (see figure \ref{fft}), hence $L_c(0)\simeq 9 L_\phi\simeq 4\mu$m, when using $L_\phi=450$nm from table 1. This could explain the observed increase in resistance beyond the WL prediction due to the proximity of strong localization. $L_\phi$ can also be estimated directly from the $B\cdot L_\phi^2$-dependence of $\Delta L_c$, noting that the width of the dip is determined by $B L_\phi^2\simeq h/e$). This yields $L_\phi\simeq 550$nm, which is consistent with $L_\phi$ obtained using McCann's theory.

We now turn to the fluctuations seen in figures \ref{WLfit} and \ref{fft}. In mesoscopic samples ($L\simeq L_\phi$) such as in exfoliated graphene, mesoscopic conductance fluctuations dominate transport properties \cite{tikh08}. Here we have a macroscopic sample where the sample is much larger than the coherence length, which averages out mesoscopic conductance fluctuations. So the remaining fluctuations are of a different nature, i.e., classical. Classical fluctuations can arise from particular disorder distributions, like ensemble averaged conductance fluctuations \cite{EACF} or as more likely here, from large scale inhomogeneities. We can estimate a lower bound on the size of the inhomogeneous regions, by assuming that the inhomogeneities are uncorrelated. For fluctuations of the order of 1\%, as seen in figures \ref{WLfit} and \ref{fft}, this would lead to $L_{corr}\gtrsim 0.01 L$, where $L\simeq 1$cm hence  $L_{corr}\gtrsim  100\mu$m. Interestingly, the typical grain size of the Cu surface used for the CVD growth is of the order of 500$\mu$m (see figure \ref{fabrication}), which we measured using electron backscatter diffraction (EBSD). Since the growth of graphene is affected by the crystal orientation, this could lead to large scale inhomogeneities with graphene grown on different grains having different densities and mobilities, which would explain the observed fluctuations of the resistance in B. These fluctuations are perfectly symmetric in B due to the small Hall resistance and are reproducible for a given contact configuration even after several weeks. However, they change drastically when changing the density by only 0.5\%, when applying a small voltage on the gate formed from the doped Si substrate as seen in fig. \ref{fft}. This also implies that it would be difficult to observe quantum oscillations at low fields. Interestingly, these large inhomogeneities can be circumvented by patterning contacts on small parts of the sample as was done in ref. \cite{cao10}, where the quantum Hall effect was observed.

Summarizing, we have studied the magnetotransport of large scale graphene monolayers obtained by CVD growth and subsequent transfer, which show many interesting features. Most importantly, they exhibit a very sharp peak at $B=0$ in resistance, which can be attributed to WL as well as strong localization. We further observed large resistance fluctuations, which might be due to the macroscopic grain boundaries from the original CVD growth and demonstrated an effective way to extract very precise Hall densities even in the presence of these large resistance fluctuations and large contact misalignments.

We would like to acknowledge A. Guermoune and M. Siaj for providing us with Si/SiO$_2$ wafers, MIAM microfab facilities and GCM at Polytechnique for processing and characterization in addition to financial support from RQMP, NSERC and FQRNT.

\end{document}